\documentclass[aps,prl,showpacs,twocolumn,floats,epsfig]{revtex4}
\usepackage{amssymb}
\usepackage{amsbsy}
\usepackage{amsmath}
\usepackage{epsfig}
\usepackage{amsfonts} 
\usepackage{bm}
\usepackage{graphicx} 
\usepackage[breaklinks,colorlinks = true,linkcolor = blue,urlcolor  = blue,citecolor = blue,anchorcolor = green,bookmarks=true]{hyperref}
\usepackage{mathrsfs}
\begin{document}

\title {Spin correlations and phase diagram of the perturbed Kitaev model}

\author{S. Mandal$^{1}$}
\author{Subhro Bhattacharjee$^{2}$}
\email{subhro@physics.utoronto.ca}
\altaffiliation[Present Address: ]{Department of Physics, University of Toronto, 60 St. George Street, Toronto, Ontario, Canada M5S 1A7.}
\author{K. Sengupta$^{3}$}
\author{R. Shankar$^{4}$}
\author{G. Baskaran$^{4}$}

\affiliation{$^{1}$International Institute of Physics, UFRN, 59078-400 Natal-RN, Brazil. \\
$^{2}$Department of Physics, Indian Institute of Science, Bangalore-560012, India.\\
$^{3}$Theoretical Physics Department, Indian Association for the
Cultivation of Science, Jadavpur, Kolkata-700032, India. \\
$^{4}$The Institute of Mathematical Sciences, C.I.T Campus,
Chennai-600113, India.}

\date{\today}

\begin{abstract}

We present a general classification of the perturbations to the
Kitaev model on the basis of their effect on it's spin correlation
functions. We derive a necessary and sufficient condition for the
spin correlators to exhibit a long ranged power-law behavior in the
presence of such perturbations. We substantiate our result by a
study of the phase diagram of the Kitaev model augmented by a loop
term and perturbed by an Ising term, within a RVB mean-field theory.
We estimate the stability of the spin-liquid phase against such
perturbations and show that this model exhibits both
confinement-deconfinement transitions from spin liquid to
antiferromagnetic/spin-chain/ferromagnetic phases as well as
topological quantum phase transitions between gapped and gapless
spin liquid phases.

\end{abstract}


\maketitle

\begin{figure}
\rotatebox{0}{\includegraphics[scale=0.45]{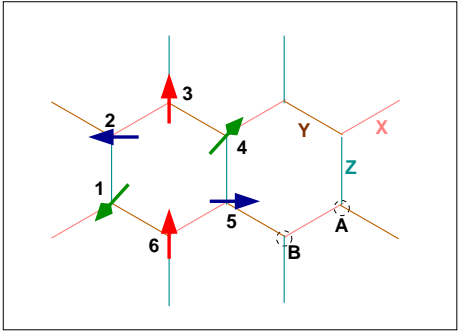}}
\caption{(Color online) Schematic representation of the Kitaev model on a honeycomb lattice showings the different links $x$, $y$ and $z$ and the two sublattices A and B. The sites labeled $1..6$ and their spin configuration is a schematic representation for a classical configuration with $\langle W_p \rangle=1$ where the green, blue and red spins point in the $x,y$ and $z$ directions respectively.}
\label{fig1}
\end{figure}

The spin-1/2 Kitaev model on a honeycomb lattice (Fig \ref{fig1}), a
rare example of an exactly solvable 2D quantum spin model, is known
to have several interesting features
\cite{kitaev1,feng1,baskaran1,lee1}. The Hamiltonian for the model
is given by
\begin{eqnarray}
H_K =\sum_{j+l={\rm even}} ( J_1 \sigma_{j,l}^x \sigma_{j+1,l}^x +
J_2 \sigma_{j-1,l}^y \sigma_{j,l}^y + J_3 \sigma_{j,l}^z
\sigma_{j,l+1}^z ), \label{ham1}
\end{eqnarray}
where $j$ and $l$ denote the column and row indices of the lattice and $\sigma^{x,y,z}$ are Pauli matrices. The model has spin liquid ground states and exhibits 
 spin fractionalization at all energy scales 
 with extremely short ranged (nearest neighbour) spin-spin correlations {\cite{baskaran1}}. The spectrum is gapless for $|J_1-J_2| \le J_3 \le J_1+J_2$ \cite{kitaev1} and gapped otherwise. The gapped phase has Abelian anyonic excitations while the gapless phase supports non-Abelian anyons which could provide topologically protected subspaces for fault tolerant quantum computation \cite{kitaev1,topocomp1}.

There have been several proposals for experimentally realizing this model in systems of ultracold atoms and molecules trapped in optical lattices, quantum circuits \cite{expt1} and in layered iridates\ \cite{jackeli}. Almost inevitably 
such realizations will have contaminating perturbations; therefore the effect of such perturbation on the spin-liquid phases of the Kitaev model deserves theoretical attention. Recently, the effect of one such perturbation, namely  a Zeeman field given by the Hamiltonian $H_z= -h \sum_i \sigma_i^z$, on the spin correlations of the model has been studied \cite{tikho1}. It was found 
that 
such a perturbation (arbitrarily weak) qualitatively alters the nature of the spin correlators by rendering them long-ranged (in power-law sense). 
However, such an analysis has not been extended to other relevant 
perturbations. In particular, there is no general theoretical criteria to classify the perturbations according to their effect on the spin correlations. Furthermore, the stability of the Kitaev spin-liquid in the presence of perturbations and the possible transitions of this perturbed Kitaev model from the deconfined spin-liquid state 
to the confined spin-ordered states 
has not been studied.

In this letter, we study both the above-mentioned issues. First, we provide an explicit classification of different short-range 
perturbations, in the gapless spin liquid phase, 
based on the nature of the spin correlators they induce.  We derive
the necessary and sufficient condition for the spin correlators to
exhibit a long ranged power-law behavior in the presence of such
perturbations and show that the induction of a long-range
spin-correlation by a magnetic field, studied in Ref.\
\cite{tikho1}, constitutes {\it a specific example} of this general
condition. Following this we take up a concrete example of the
second class of perturbation which does not induce long-range
correlations and study the Kitaev model in the presence of a loop
term and the Ising-like perturbative term . We analyze this model using
a RVB mean-field approach and demonstrate that the model exhibits transitions from a deconfined spin liquid to confined Ising ordered antiferromagnetic (AFM), ferromagnetic (FM) or spin-chain (SC) phases. 
We also show that the model supports two distinct spin liquid phases with gapped and gapless deconfined spinon excitations 
with a topological quantum phase transition separating 
them. We chart out the phase diagram of this model and estimate the stability of the deconfined phases as a
function of the strengths of the loop and Ising terms. 
Both the issues studied here are highly experimentally relevant and 
our work 
constitutes a significant extension of our understanding of the perturbed Kitaev model.

We begin with the study of the fate of spin correlations of the Kitaev model under most general class of short ranged perturbations. The total Hamiltonian is then given by $H= H_K+H_P$, where $H_P$ is a perturbing Hamiltonian made up of spin operators. Examples of $H_P$ include $H_z$ or the most general bilinear spin-spin interaction term given by $H_b=\sum_{\langle ij\rangle} \lambda_{\alpha\beta}\sigma^\alpha_i\sigma^\beta_j$ including the experimentally most relevant anisotropic Heisenberg interaction $H_h= \sum_{\langle ij\rangle}\left(\lambda_{xx}\sigma^x_i\sigma^x_j +\lambda_{yy}\sigma^y_i\sigma^y_j+\lambda_{zz}\sigma^z_i\sigma^z_j\right)$.

\begin{figure}
\rotatebox{0}{\includegraphics*[scale=0.2]{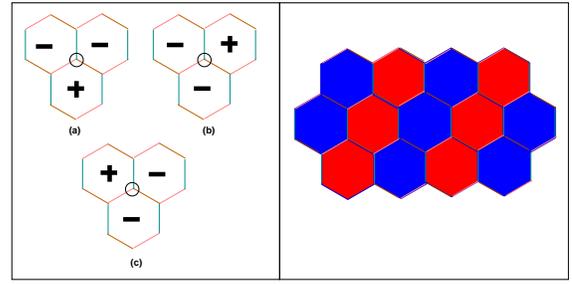}}
\caption{(color online) Left Panel: The flux flipped on application of $\sigma_i^\alpha$ where $i$ is the site at the center (circled): (a) $\alpha=z$, (b) $\alpha=y$ (c) $\alpha=x$ (see Fig. \ref{fig1} for definition of the $x,y,z$ bonds).  Right Panel: A schematic representation of $y$ type covering $\Gamma_{yR}$ (red plaquettes) and $\Gamma_{yB}$ (blue plaquettes). A product of $W_p$ over red or blue plaquettes 
yields $\Sigma^y$. Analogous coverings for $x$ and $z$ type coverings may be obtained by rotating the above figure by $\pm 60^{\circ}$}
\label{fluxcreate}
\end{figure}

In the analysis of the effects of these perturbations a crucial role
is played by the $Z_2$ flux operators defined as $W_p= \sigma_1^x
\sigma_2^y \sigma_3^z \sigma_4^x \sigma_5^y \sigma_6^z$ (Fig.\
\ref{fig1}) on every plaquette. Since $[H_K,W_p]=0$ these fluxes are
frozen in the unperturbed model and this fact is crucial to the
integrability of the unperturbed Kitaev Model. For the ground state
of $H_K$, $W_p=1$ $\forall p$. However, $H_P$ does not commute with
the $W_p$s and thus the $Z_2$ fluxes acquires dynamics in the
perturbed model. The effect of this dynamics is key to determining
the nature of the spin correlators of the perturbed Kitaev model. To
this end we define 3 conserved operators,
$\Sigma^{\alpha}\equiv\prod_j(i\sigma_j^{\alpha}) = \prod_{p \in
\Gamma_{\alpha c}} W_p$ where $\alpha=x,y,z$ and the product over
$W_p$s in the second expression is taken over a subset of plaquettes
$\Gamma_{\alpha c}$ in the following way. Colour all the plaquettes
with two colours, red and blue, as shown in Fig.\ {\ref{fluxcreate}}
(right panel). Then $\Gamma_{yR}(\Gamma_{yB})$ denotes product of
all $W_p$s on the red(blue) plaquettes which gives $\Sigma^y$.
Analogous coverings in the other two directions yields $\Sigma^x$
and $\Sigma^z$. The $\Sigma^\alpha$ operators correspond to global
$\pi$ spin-rotations about the $\alpha$th axis and with periodic
boundary conditions we have $\Sigma^x\Sigma^y =(-1)^N \Sigma^z$
where $N$ is the total number of sites.

With these definitions, we now state and prove the necessary and
sufficient condition that the perturbation $H_P$ {\em does not}
change at least one dynamic spin-spin correlation ({\em i.e.}
$(\langle\langle
\sigma_r^\alpha(t)\sigma^\alpha_0(0)\rangle\rangle)$ ) from short
ranged to a long ranged one (in power-law sense). The condition is:
\begin{eqnarray}
[\Sigma^{\alpha},H_P]_{-}= 0\ \ \ \ \forall \alpha,\label{cond1}
\end{eqnarray}
where $[..]_{-(+)}$ denotes the commutator (anti-commutator). {\em
If this condition is violated, then at least one component of the
dynamical spin-spin correlation becomes a power law}. To prove this
we take the case of $\langle\langle
\sigma_r^z(t)\sigma_0^z(0)\rangle\rangle$ for concreteness
(\textcolor{black}{without loss of generality}). We note that in
terms of the Majorana Fermions we have
$\sigma_i^\alpha=-ib^\alpha_ic_i$ \cite{baskaran1,feng1,kitaev1}.
Thus the operation of $\sigma_i^\alpha$ creates a $c$ Fermion on the
$i^{th}$ site as well as two quanta of $Z_2$ flux on adjacent
plaquettes which share an $\alpha$-type bond. This is schematically
illustrated in Fig.\ \ref{fluxcreate} (left panel). Thus the
operators $\sigma_r^\alpha$ and $\sigma_0^\alpha$ create such fluxes
centered around $r$ and $0$ respectively. Since the perturbation
does not conserve flux, one can always repeatedly apply $H_P$ and
move the pair of flux centered at $r$ to the pair centered at $0$
and then annihilate them. This occurs at order $\sim r$ of the
perturbation theory and one has a string of perturbation operators
joining $0$ and $r$. Such a process has a leading term
$\lambda^r=e^{-r\ln{(1/\lambda)}}$ which gives rise to exponential
decay of spin correlations with correlation length $\xi\sim
1/\ln{(1/\lambda)}$.

The above discussion indicates that in order to get long-ranged
correlations, the perturbation term must neutralize the two pairs of
fluxes centered at $r$ and $0$ locally and independently without
forming the intermediate string whose length scales with $r$. This
requires that the application of $\sigma_i^\alpha$ changes
$\Sigma^\beta$ for at least two values of $\beta$, {\em e.g.}
$[\Sigma^{x(y)},\sigma_i^z]_+ = 0$ and $[\Sigma^z,\sigma_i^z]_-=0$.
Thus the perturbing Hamiltonian, $H_P$, must be able to change the
corresponding $\Sigma^a$s in order to neutralize the flux locally.
This is possible when the generic condition
\begin{eqnarray}
[\Sigma^{\beta},H_P]_+=0
\label{cond2}
\end{eqnarray}
is fulfilled {\em for the corresponding two pair of $\beta$s}. Eqs.\
\ref{cond1} (\ref{cond2}) represents the necessary and sufficient
conditions that a perturbing Hamiltonian $H_P$ does not (does)
change the nature of the spin correlators.

To illustrate these conditions via specific examples, we consider $H_p=H_z$ (studied in Ref.\ \cite{tikho1}). It is easy to see that $[\sigma^z_r,\Sigma^{x(y)}]_+=0$ and $[\sigma_r^z,\Sigma^z]_-=0$. Thus the $Z_2$ fluxes are neutralized locally leading to long-ranged spin correlators. 
For a Zeeman term in the $z$ direction, 
only the $zz$ correlators become power-law while $xx$ or $yy$ correlators remain exponential as before. The above results hold for $H_P=H_b$ when $\alpha\neq\beta$. Next, we consider $H_P=H_h$. The individual terms in $H_h$ are of the form $\sigma_i^\alpha\sigma_j^\alpha$, where $i$ and $j$ are nearest neighbors. If $\langle ij\rangle$ is in the $\alpha^{\rm th}$ direction, they commute with all $W_p$. Otherwise, each spin operator flips 
two distinct pairs of $W_p$s on the plaquettes which share the bond
$i,i+e_a$ and $j,j+e_a$. However, it is easy to check that
$\Sigma^\alpha$ contains either both or none of the flipped $W_p$s
and hence  $H_h$ commutes with all the three $\Sigma^{\alpha}$.
$H_h$ therefore fails to destroy the short ranged nature of the
spin-spin correlations.

Having established the general setting, let us now look at the spin-spin correlation in detail. A typical $n^{\rm th}$ order term in the perturbation expansion for the spin-spin correlation is, $T_n = \sum_{\langle i_1j_1\rangle}\cdots\sum_{\langle i_nj_n\rangle}\int d\tau_1\cdots\int d\tau_n\langle{\cal T}\Big (\sigma^\alpha_r(\tau)h^{a_1}_{i_1j_1}(\tau_1)\\ h^{a_2}_{i_2j_2}(\tau_2)...h_{i_nj_n}(\tau_n)\sigma^\alpha_0(0)\Big)\rangle/n!$, where $h^a_{ij}$ stand for the individual terms in $H_P$. Here we have taken all operators to evolve in Euclidean time, $O(\tau)=e^{H_K\tau}Oe^{-H_K\tau}$. Consider now the case when $r$ is sufficiently far from the origin. At any finite order, $n$ (with $n$ being even) $T_n$ may be written as $T_n = \sum_{\langle i_1j_1\rangle}\cdots\sum_{\langle i_nj_n\rangle}\int d\tau_1\cdots\int d\tau_n\langle{\cal T}\Big ([\sigma^\alpha_r(\tau) h^{a_1}_{i_1j_1}(\tau_1)\\ h^{a_2}_{i_2j_2}(\tau_2)...]\left[\cdots h_{i_nj_n}(\tau_n)\sigma^\alpha_0(0)\right]\Big)\rangle/n!$, where we have divided up the series using $[\cdots]$ such that the fluxes are neutralized around $r$ and $0$ respectively by the operators belonging to each group within the square brackets. Once the fluxes are neutralized, the $b_i^\alpha$ Fermions are no longer important except for an overall constant. 
The $c$ Fermions now determine the details of the correlation. In
the gapless phase of the Kitaev model the $c$ fermions have a
Fermi-surface and gapless excitations across it. Thus the
$c$-Fermion propagator is an $n^{\rm th}$ order free Fermion
propagator, which in $(2+1)$ dimension, is given by
$\mathcal{G}_c(\beta,r,\tau)\sim
\left(r^2+\tau^2\right)^{-(np/2+1)}$, where $p$ is the number of
$\sigma$ operators occurring in $h_{ij}^a$. Thus we expect that the
connected spin-spin correlation function goes as:
\begin{eqnarray}
\langle\langle\sigma_r^\alpha(t)\sigma_0^\alpha(0)\rangle\rangle
\sim \lambda^n{\left(r^2-t^2\right)^{-(np/2+1)}},
\label{corrform}
\end{eqnarray}
where $\lambda$ is the coupling constant. 
Eq.\ \ref{corrform} reproduces the results of Ref.\ \cite{tikho1} where $p=1$ and $n=2$. 

We note that the
above results are valid for infinite 2D systems and our conclusion may change for finite systems such as nanotubes, {\em i.e.}, a cylinder of infinite length and (finite) perimeter $L$. 
In this case it may be possible to annihilate the $Z_2$ flux locally by going around the finite direction of the
cylinder even for a perturbation which does not induce power-law correlations in an infinite system. This leads to a crossover of the behavior in the correlation function. For $L\gg r$, it is easier to construct the string joining $0$ and $r$ (at the lower order of perturbation theory) than going around the cylinder and the infinite geometry results hold. However, for $r \ll L$, it is easier to form a string round the axis of the cylinder 
annihilation of the $Z_2$ flux leading to to power-law correlators. Such a term occurs at order $\sim \lambda^L$ of the perturbation theory.

Next, we study the phase diagram of the Kitaev model augmented by a loop term $H_L= -\kappa \sum_p W_p$ and perturbed by an Ising-like perturbation $H_I= \lambda \sum_{\langle ij\rangle} S_i^z S_j^z $ for $J_{1,2,3}=1$ and within a RVB mean-field theory. To this end, we use our earlier transformation $\sigma_i^\alpha=-ib^\alpha_ic_i$ to map the spin model $H_K+H_L+H_I$ to a Fermionic model $H_F$. 
The resultant Hamiltonian becomes
\begin{eqnarray}
H_F &=& - \sum_{j\in A} \Big[ \sum_{\alpha=x,y\, links } i c_{j}
c_{j_{\alpha}^{'}} + \sum_{z \,link}i b^z_{j} c_j i b^z_{j_z^{'}}
c_{j_z^{'}} \Big]
\nonumber \\
&& -\kappa \sum_{j,k \in plaquette} \sum_{z link} i b_j^z
b_{j_z^{'}}^z i b_k^z b_{k_z^{'}}^z \nonumber\\
&& + \lambda \sum_j \sum_{\alpha = all \,links} i b^z_{j} c_j i
b^z_{j_{\alpha}^{'}} c_{j_{\alpha}^{'}},
\label{maj1}
\end{eqnarray}
where the subscript $j,k \in plaquette$ indicates that the sum is over sites which belong to the $A$ sublattice of a given plaquette as schematically shown in Fig.\ \ref{fig1}. Note that for $\lambda=0$, the operators $ib^z_jb^z_{j^\prime_z}$, commute with the Hamiltonian and are therefore a constants of motion. In this limit, $H$ is exactly solvable. When $\lambda$ is turned on, these operators acquire dynamics and their fluctuations are ultimately expected to confine the spinons through a confinement-deconfinement transition.

To make further progress, we introduce RVB type mean-fields \cite{baskaran1,lee1} on the sites (corresponding to spin ordering) and on links (corresponding to the emergent gauge fields) of the hexagonal lattice: $\langle i b_j^z c_j\rangle = \langle \sigma^z_j \rangle = \Delta_{1(2)}$, $\langle i b^z_j c_{j_{\alpha}^{'}}\rangle = \beta_{\alpha}$, $\langle i b^z_j b^z_{j_{\alpha}^{'}}\rangle = \gamma_{\alpha}$, and $\langle i c_j c_{j_{\alpha}^{'}}\rangle = \gamma_{0\alpha}$. Note that keeping in mind the bipartite nature of the hexagonal lattice and to allow for possible AFM phases, we have introduced two mean-fields $\Delta_1$ and $\Delta_2$ corresponding to the two sublattices shown in Fig.\ \ref{fig1}. 
Decomposing the quartic term in $H_F$ in using these mean fields we have 
the quadratic mean-field Hamiltonian, which, in momentum space, is given by 
\begin{eqnarray}
H_{\rm mf} &=& \frac{1}{N} \sum_{\vec k} \Big[ J_0 \left( \alpha +
e^{i k_1} +
e^{i(k_1+k_2)} \right) c_{\vec k}^{A \dagger} c_{\vec k}^{B} \nonumber\\
&& +J_0^{'} \left( \beta -2 \kappa \gamma_z /J_0^{'} + e^{i k_1} +
e^{i
(k_1+k_2)} \right) b_{\vec k}^{A \dagger} b_{\vec k}^{B} \nonumber\\
&& + (i c_{\vec k}^{A \dagger} b_{\vec k}^{B}- i b_{\vec k}^{A
\dagger} c_{\vec k}^{B} ) \left( \beta_z (1+\lambda) + \beta_x
e^{ik_1} \right.
\nonumber\\
&& \left. + \beta_y e^{i k_2} \right)- c_1 b_{\vec k}^{A \dagger}
c_{\vec k}^{A} - c_2 b_{\vec k}^{B
\dagger} c_{\vec k}^{B} + \rm {h.c} \Big]  \nonumber\\
&& + \kappa \gamma_z^2 - (1+\lambda) \gamma_z \gamma_{0z} + (1+3
\lambda) \Delta_1 \Delta_2,
 \label{mf2}
\end{eqnarray}
\begin{figure}
 \rotatebox{0}{\includegraphics*[width=.45\linewidth]{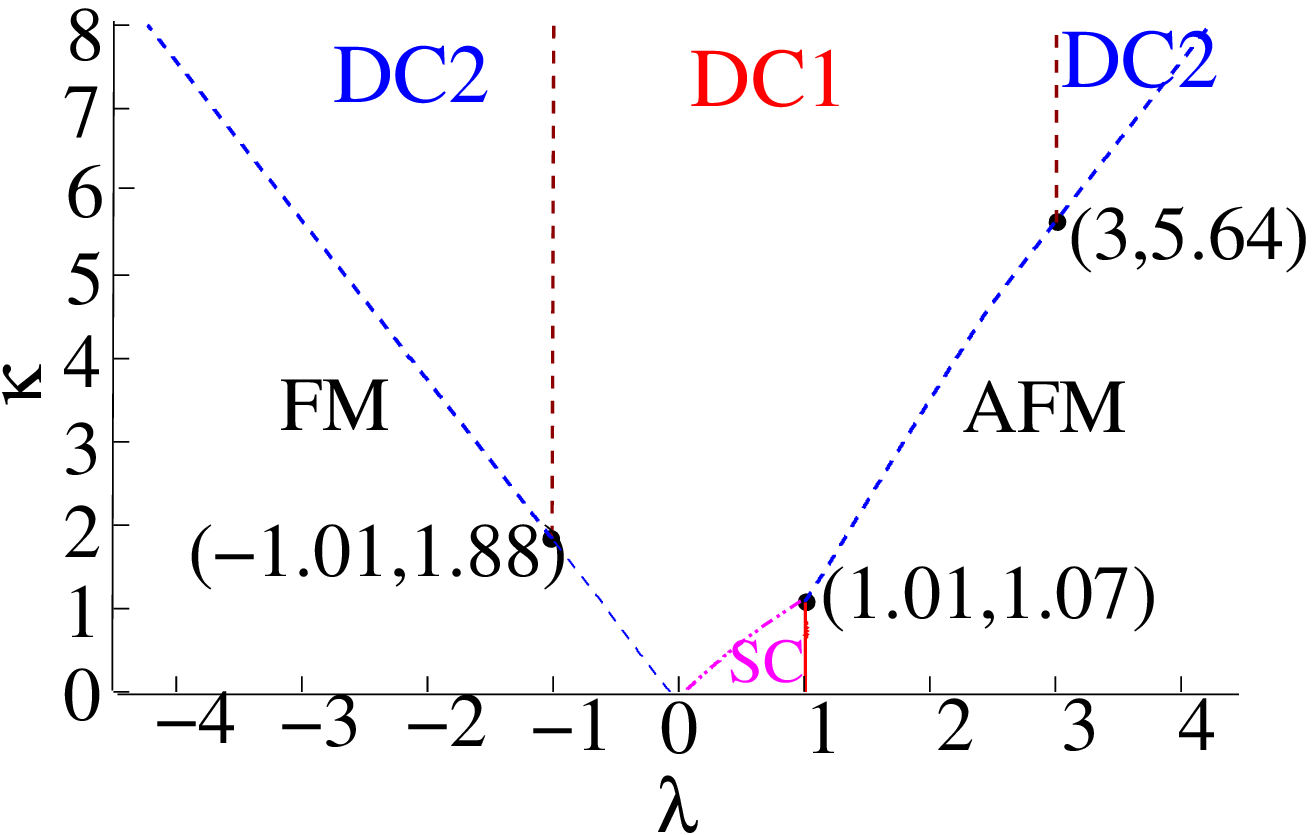}}
 \rotatebox{0}{\includegraphics*[width=.45\linewidth]{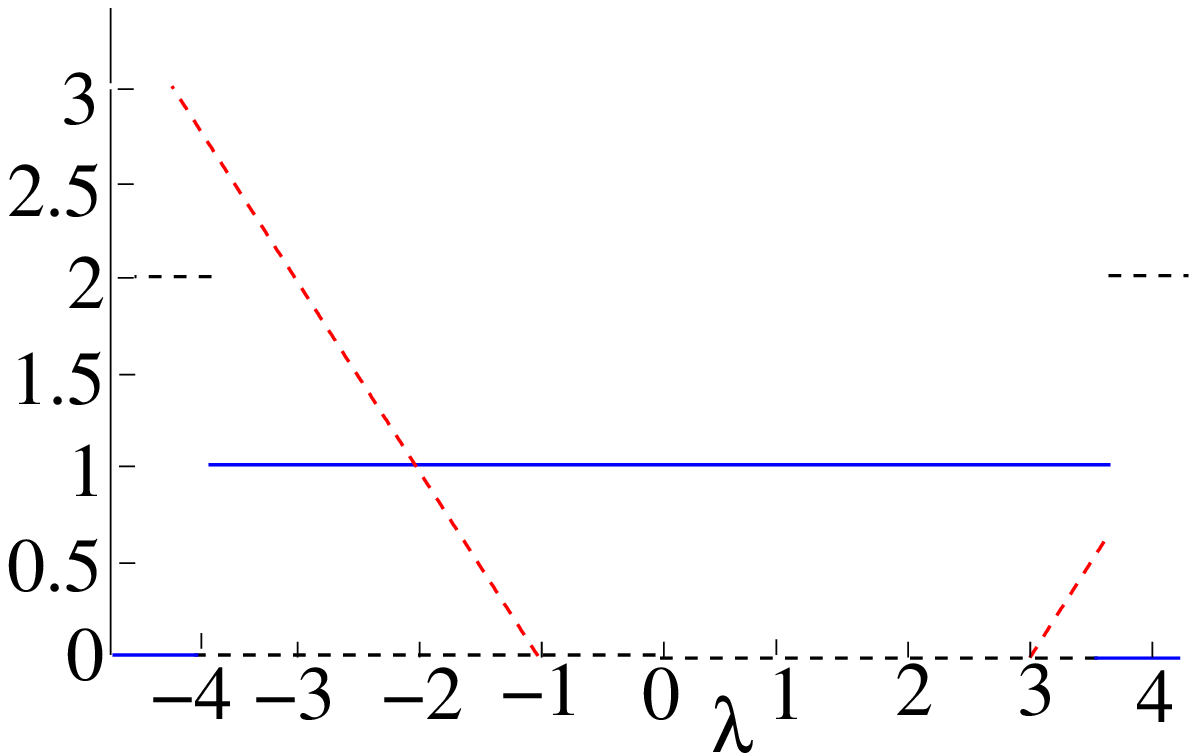}}
\caption{(Color online) Left Panel: The mean-field phase diagram for
the model. The blue dashed lines represent confinement-deconfinement
transitions. 
The triple points occur at
$(\lambda^{\ast}_1,\kappa^{\ast}_1)=(1.01,1.07)$ and
$(\lambda^{\ast}_2,\kappa^{\ast}_2)=(3,5.64)$ for $\lambda>0$ and
$(\lambda^{\ast}_3,\kappa^{\ast}_3)=(-1.01,1.88)$ for $\lambda <0$.
Right Panel: Plot of the loop order parameter (solid blue line), the
spinon gap (red dashed line) and the FM and the AFM order parameters
(black dashed lines) as a function of $\lambda$ for $\kappa=7$.
} \label{fig2}
\end{figure}
where $J_0= (1+\lambda \gamma_{x})$, $\alpha J_0 =
\left(1+\lambda\right) \gamma_z$, $J_0^{'}= \lambda \gamma_{0x}$,
$\beta J^{'}_0 = \left(1+\lambda\right) \gamma_{0z}$, $c_{1(2)}=
(1+3\lambda) \Delta_{1(2)}$, and the momentum $\vec k = k_1 \hat e_1
+ k_2 \hat e_2$ with the unit vectors $\hat e_1 = \hat x + \hat
y/\sqrt{3}$ and $\hat e_2 = 2\hat y/\sqrt{3}$.

We now minimize $H_{\rm mf}$ numerically and obtain the mean-field
phase diagram of the model as a function of $\lambda$ and $\kappa$.
Note that the mean field solution is exact at $\lambda=0$. This
phase diagram is shown in the left panel of Fig.\ \ref{fig2}. In
accordance with our expectation, we find that at large positive
(negative) $\lambda$, the ground state of the model is an Ising AFM
(FM) which corresponds to confined phase of spinons while at small
$\lambda$, the model exhibits a deconfined gapless phase DC1. The
transition between DC1 and AFM , at low $\kappa$ and positive
$\lambda$, occurs via a intermediate SC phase, which corresponds to
antiferromagnetic alignment of spins along chains in $x$ direction
of the hexagonal lattice with ferromagnetic arrangement of such
chains in the $y$ direction. For negative $\lambda$, there is a
direct transition to the FM phase from DC1 (for small $\kappa$). At
high enough values of $\kappa$, we find another gapped deconfined
phase DC2. The transition between DC1 and DC2 is and second order
within mean field theory and is an example of a topological quantum
phase transition. The confinement-deconfinement transitions at high
$\kappa$ always occur from DC2 to AFM/FM phases. These transitions
are predicted to be first order within mean-field theory. The phase
diagram exhibits two triple points at $(\lambda^{\ast}_1,
\kappa^{\ast}_1)=(1.01,1.07)$ and $(\lambda^{\ast}_2,
\kappa^{\ast}_2)=(3,5.64)$ for $\lambda > 0$. These represent
meeting points of AFM, SC and DC1 and AFM, DC2 and DC1 phases
respectively. For $\lambda <0$, there is one triple point
$(\lambda^{\ast}_3, \kappa^{\ast}_3)=(-1.01,1.88)$ where the FM,
DC1, and DC2 phases meet. We also note that our mean-field analysis
also gives an estimate for the stability of the deconfined phase of
the Kitaev model ($-0.07 \le \lambda_c \le 0.08$ for $\kappa=0$)
under external perturbing Ising term which may be important for
physical realization of the Kitaev model and for quantum computing
proposals based on it \cite{expt1,topocomp1}.

The plot of the loop order parameter $\langle W_p \rangle$, the
spinon gap, and the AFM and the FM order parameters as obtained from
the mean-field theory, is shown, for $\kappa=7$, as a function of
$\lambda$ in the right panel Fig.\ \ref{fig2}. We note that all the
order parameters show discontinuous changes at the transition points
indicating first order transitions. The spinon gap, in contrast,
increases linearly and continually with $\lambda$ indicating a
second order quantum phase transition between DC1 and DC2 phases.
The presence of this topological quantum phase transition and the
linear variation of the spinon gap with $\lambda$ can be understood
qualitatively from $H_{\rm mf}$. For large $\kappa$, it requires a
large $\lambda$ to destabilize the Kitaev ground state in favor of
Ising AFM/FM. In addition, numerically we find that in the Kitaev
phase $\gamma_z (\gamma_x) \sim 1(0)$. As a result, beyond a
critical value of $\lambda=\lambda_c$, the effective couplings along
the links, $J_{1,2} \sim (1+ \lambda \gamma_x)$, $J_3 \sim
\gamma_z(1+\lambda)$, fail to satisfy $|J_1-J_2|\le J_3 \le J_1+J_2$
thus leading to a gapped phase via a topological quantum phase
transition \cite{kitaev1,feng1}. The spinon gap in this gapped phase
varies linearly with $J_3$ \cite{feng1,baskaran1} and hence shows a
linear variation on $\lambda$. At small $\kappa$, the
confinement-deconfinement transitions to the SC/FM phases occur
before $\lambda_c$ is reached and hence the topological phase
transitions do not occur.

To conclude, we have presented a general classification of the contaminating interactions
of the Kitaev model based on their effects on the spin correlators of the model and have
derived a necessary and sufficient condition for the interaction to induce power-law
spin-spin correlations. We have also presented the phase diagram of the Kitaev model,
augmented by a loop term and perturbed by an Ising Hamiltonian, and have shown that the
model exhibits a rich phase diagram with several interesting transitions.
Our estimate suggests that the topological phase of the Kitaev model
is unstable to about 10\% contamination by Ising interactions.

SB thanks  R. Moessner for valuable discussions and acknowledges
hospitality of IMSc where a part of this work was done. KS thanks
DST, India for support through grant SR/S2/CMP-001/2009.

\end{document}